\documentclass[conference,10pt]{IEEEtran}
\normalsize
\usepackage{bbm}
\usepackage{adjustbox}
\usepackage{enumitem}
\usepackage{cite,algorithm,algorithmic,amsmath,amssymb,amsthm,empheq,mhsetup}
\usepackage{subfigure,amsfonts,balance}
\usepackage{epstopdf}
\usepackage{enumitem}
\usepackage{setspace}
\usepackage[dvipsnames]{xcolor}
\usepackage[left=0.65in, right=0.65in, top=0.7in, bottom=0.96in,centering]{geometry}
\DeclareMathOperator{\tr}{\mathrm{tr}}

\DeclareMathOperator{\EEE}{\mathbb{E}}

\DeclareMathOperator{\C}{\mathbb{C}}

\DeclareMathOperator{\aaa}{\pmb{a}}

\DeclareMathOperator{\K}{\mathcal{K}}

\DeclareMathOperator{\z}{\pmb{z}}
\DeclareMathOperator{\x}{\pmb{x}}

\DeclareMathOperator{\HH}{\pmb{H}}

\DeclareMathOperator{\LL}{\mathcal{L}}

\DeclareMathOperator{\CN}{\mathcal{CN}}

\DeclareMathOperator{\NN}{\mathcal{N}}
\DeclareMathOperator{\CC}{\pmb{C}}

\DeclareMathOperator{\W}{\pmb{W}}

\DeclareMathOperator{\y}{\pmb{y}}

\DeclareMathOperator{\uu}{\pmb{u}}

\DeclareMathOperator{\g}{\pmb{g}}

\DeclareMathOperator{\X}{\pmb{X}}
\DeclareMathOperator{\Y}{\pmb{Y}}

\DeclareMathOperator{\UPSI}{\pmb{\upsilon}}
\DeclareMathOperator{\PHI}{\pmb{\Phi}}

\DeclareMathOperator{\VARPHI}{\pmb{\varphi}}

\DeclareMathOperator{\diag}{\mathrm{diag}}

\newtheorem{proposition}{Proposition}

\ifCLASSINFOpdf
\else
\fi

\begin{document}
	\bstctlcite{IEEEexample:BSTcontrol}
	\fontsize{10}{12}\rm
	%
	% paper title
	% can use linebreaks \\ within to get better formatting as desired
	\title{\Huge Channel Estimation in RIS-assisted Downlink Massive MIMO: A Learning-Based Approach \vspace{-2mm}}
	\author{
		\IEEEauthorblockN{
			Tung T. Vu\IEEEauthorrefmark{1},
			Trinh Van Chien\IEEEauthorrefmark{2},
			Canh T. Dinh\IEEEauthorrefmark{3},
			Hien Quoc Ngo\IEEEauthorrefmark{1}, and
			Michail Matthaiou\IEEEauthorrefmark{1}
			%Duy T. Ngo\IEEEauthorrefmark{1},
			%Minh N. Dao\IEEEauthorrefmark{3}
			%Nguyen H. Tran\IEEEauthorrefmark{4},
			%and Richard H. Middleton\IEEEauthorrefmark{1}
			%Minh N. Dao\IEEEauthorrefmark{1},
			%Salman Durrani\IEEEauthorrefmark{4},
			%and Richard H. Middleton\IEEEauthorrefmark{1}
		}
		\IEEEauthorblockA{\small\IEEEauthorrefmark{1}Centre for Wireless Innovation (CWI), Queen's University Belfast, Belfast BT3 9DT, UK}
		\IEEEauthorblockA{\small\IEEEauthorrefmark{2}School of Information and Communication Technology (SoICT), Hanoi University of Science and Technology, Vietnam}
		%\IEEEauthorblockA{\small\IEEEauthorrefmark{4}School of Computer Science, The University of Sydney, Sydney, NSW 2006, Australia}
		\IEEEauthorblockA{\small\IEEEauthorrefmark{3}School of Computer
			Science, The University of Sydney, Sydney, NSW 2006, Australia}
		\IEEEauthorblockA{
			%\small\IEEEauthorrefmark{3}School of Engineering, Information Technology and Physical Sciences, Federation University Australia, Ballarat, VIC 3353, Australia
			%\\
			Email: t.vu@qub.ac.uk, chientv@soict.hust.edu.vn, canh.dinh@sydney.edu.au, hien.ngo@qub.ac.uk, m.matthaiou@qub.ac.uk
			%, duy.ngo@newcastle.edu.au, m.dao@federation.edu.au
			%, nguyen.tran@sydney.edu.au, richard.middleton@newcastle.edu.au
		}
		%\author{
			%\IEEEauthorblockN{Author 1 and Author 2}
			%\IEEEauthorblockA{\small\\
				% \\
				%    }
			%\thanks{This work is supported in part by an ECR-HDR scholarship from The University of Newcastle and in part by an Australian Research Council Discovery Project grant DP170100939.}
			\vspace{-9mm}
		}
		
		\maketitle
		\allowdisplaybreaks
		\vspace{-0mm}
		\begin{spacing}{1}
			\begin{abstract}
				For downlink massive multiple-input multiple-output (MIMO) operating in time-division duplex protocol, users can decode the signals effectively by only utilizing the channel statistics as long as channel hardening holds. However, in a reconfigurable intelligent surface (RIS)-assisted massive MIMO system, the propagation channels may be less hardened due to the extra random fluctuations of the effective channel gains. To address this issue, we propose a learning-based method that trains a neural network to learn a mapping between the received downlink signal and the effective channel gains.  The proposed method does not require any downlink pilots and statistical information of interfering users. Numerical results show that, in terms of  mean-square error of the channel estimation, our proposed learning-based method outperforms the state-of-the-art methods, especially when the light-of-sight (LoS) paths are dominated by non-LoS paths with a low level of channel hardening, e.g., in the cases of small numbers of RIS elements and/or base station antennas.
			\end{abstract}
		\end{spacing}
		
		% IEEEtran.cls defaults to using nonbold math in the Abstract.
		% This preserves the distinction between vectors and scalars. However,
		% if the journal you are submitting to favors bold math in the abstract,
		% then you can use LaTeX's standard command \boldmath at the very start
		% of the abstract to achieve this. Many IEEE journals frown on math
		% in the abstract anyway.
		% Note that keywords are not normally used for peerreview papers.
		%\vspace{-0mm}

		% For peer review papers, you can put extra information on the cover
		% page as needed:
		% \ifCLASSOPTIONpeerreview
		% \begin{center} \bfseries EDICS Category: 3-BBND \end{center}
		% \fi
		%
		% For peerreview papers, this IEEEtran command inserts a page break and
		% creates the second title. It will be ignored for other modes.
		\IEEEpeerreviewmaketitle
		
		%\balance
		\vspace{-3mm}
		\section{Introduction}
		\vspace{-1mm}
		\label{sec:Introd}
		Reconfigurable intelligent surfaces (RIS) have currently been considered as an emerging technology to enhance the coverage and capacity of wireless communication systems with low hardware cost and energy consumption \cite{wei21CL}. By reconfiguring the low-cost meta-material elements of an RIS to create a phase-shift pattern over the surface that reflects an incident wave of a beam in a specific direction, the radio environments can be controlled to boost signal power at the receivers even at unfavorable locations \cite{chien21TWC}. However, RISs make the channel models from the transmitter and receiver more complicated than those in the systems without RISs, making channel estimation a key challenge. Therefore, it is essential to develop efficient channel estimation approaches to fully exploit the potentials of RIS-assisted communication systems. 
		
		Massive multiple-input multiple-output (MIMO) is the key technology used in 5G systems that serves simultaneously users (UEs) on the same frequency band \cite{chien21TWC}. Although channel estimation has been widely studied in the traditional massive MIMO space, there are only several works that consider channel estimation in RIS-assisted massive MIMO systems, such as \cite{demir21arxiv,zhang20SPAWC,ahmet21arxiv} and references therein. In \cite{demir21arxiv}, a channel estimation scheme was proposed by exploiting the spatial correlation characteristics at both the base station (BS) and RISs. The work of \cite{zhang20SPAWC} proposed a least square-based channel estimation approach in a scenario of one single UE. In \cite{ahmet21arxiv}, the authors developed a two-stage approach using on/off RIS elements and federated learning for channel estimation. However, these existing works  focus only on the channel estimation at the BS. 
		
		\textit{Paper contributions}: In this work, we consider a RIS-assisted downlink massive MIMO system. For the downlink transmission, the BS first estimates the aggregated channels based on its received uplink pilot signals. A closed-form expression for the minimum mean square error (MMSE) channel estimates is derived. Then, the BS uses these channel estimates to precode the symbols before sending them to the UEs. At the UEs, to detect their desired signals, the UEs need to estimate the effective channel gains.
		To do this, we introduce two approaches, namely \textit{Hardening Bound} and\textit{ Model-Based} schemes that extend the state-of-the-art approaches from the traditional massive MIMO systems to the RIS-assisted massive MIMO systems exploiting the channel hardening characteristics of massive MIMO. However, these two approaches might not perform well in the RIS-assisted massive MIMO when the channels are less hardened with extra random fluctuations. Therefore, we propose a novel learning-based approach that uses deep neural networks (DNN) at the UEs to estimate their effective channels. The DNN model of the learning-based approach is trained in an offline mode with the input features obtained from the Hardening Bound and Model-based approaches. The proposed learning-based approach does not require any downlink pilots and  statistical information of interfering UEs. We validate in the simulation results that the proposed learning-based approach outperforms the Hardening Bound and Model-Based approaches, especially when the light-of-sight (LoS) paths are dominated by non-LoS paths with a low level of channel hardening (e.g., due to a small number of RIS elements and/or BS antennas).
		
		% \emph{Notation:}
		% We use boldface symbols for vectors and capitalized boldface symbols for matrices.
		% $\RRR^d$ denotes a space where its elements are real vectors of length $d$.
		% % $\rnd(x)$ is the nearest integer of $x$.
		% %$\langle\x,\y\rangle$ means the inner product of vectors $\x$ and $\y$.
		% $\pmb{X}^*$ and $\pmb{X}^H$ represent the conjugate and conjugate transpose of a matrix $\pmb{X}$, respectively.
		% %$\pmb{I}$ and $\pmb{0}$ are the identity and zero matrices with appropriate dimensions, respectively.
		% %the real part of a complex number $x$ is denoted as $\mathfrak{R}\{x\}$.
		% %For a scalar $x$, $\lfloor x\rfloor$ denotes the largest integer that is not larger than $x$.
		% %$\pmb{X}$ and $\pmb{x}$ are respectively denoted as a complex matrix and a vector.
		% %$||.||$  denotes the $\ell_2$-norm function.
		% %$||.||_0$ and $\one_{\{.\}}$ denotes the $\ell_0$-norm and the indicator function, respectively.
		% $\CN(\pmb{0},\pmb{Q})$ denotes the circularly symmetric complex Gaussian distribution with zero mean and covariance $\pmb{Q}$.
		% % while $\NN(0,V)$ the normal distribution with zero mean and variance $V$. 
		% %$\nabla g$ is the gradient of a function $g$.
		% %$|\GG|$ stands for the number of elements in set $\GG$.
		% $\EEE\{x\}$ denotes the expected value of a random variable $x$.
		
		% \begin{figure}[!t]
			%  \centering
			%  {\includegraphics[width=0.4\textwidth]{RISmMIMO.eps}\label{fig:a}}
			%  \caption{An RIS-assisted downlink massive MIMO system.}
			%  \label{Fig:systmod}
			%  \vspace{-0mm}
			% \end{figure}
		
		\vspace{-2mm}
		\section{System Model}
		\vspace{-1mm}
		\label{sec:SystemModel}
		We consider the downlink of an RIS-assisted massive MIMO system, where an $M$-antenna base station (BS) serves $K$ single-antenna UEs in the same time-frequency resource. The system operates under the time-division duplexing (TDD) scheme that includes two phases: (i) Uplink training for channel estimation, and (ii) Downlink data transmission. Since the UEs are far from the BS, the downlink transmission is assisted by $L$ RISs, each of which comprises $N$ engineered reflecting elements that can modify the phase shift of the incident signals. The modification of the phase shifts is realized by RIS controllers that exchange information, e.g., channel state information (CSI) and RIS phase shifts, with the BS via backhaul links. 
		% (see Fig.~\ref{Fig:systmod}).
		Let $\K \triangleq \{1,\dots,K\}$, $\NN \triangleq \{1,\dots,N\}$, and $\LL \triangleq \{1,\dots,L\}$. 
		Denote by $\PHI_\ell\triangleq \diag([
		% e^{j\theta_{\ell,1}},\dots,e^{j\theta_{\ell,N}}
		\nu_{\ell,1},\dots,\nu_{\ell,N}
		]) \in \C^{N \times N}$ the matrix of the phase shifts of RIS $\ell$, where $\nu_{\ell,n}$ is the phase shift applied at element $n$ of RIS $\ell$ and $n\in\NN$, $\diag(\x)$ is the diagonal matrix whose elements in the diagonal are the elements of a vector $\x$. We assume that $\PHI_\ell, \forall\ell$, are known at the BS. 
		% , where $\theta_{\ell,n} \in [-\pi,\pi]$ is the phase shift applied at element $n$ of RIS $\ell$. 

		\vspace{-2mm}
		\subsubsection{Channel Model}
		The wireless channels are modeled by the standard block-fading model \cite[Sec. 2]{ngo16}, where the time-frequency resources are divided into coherence intervals. In each coherence interval, the channels are relatively static and frequency flat. The number of symbols that can be transmitted in each coherence interval is $\tau_c$. 
		Let $\g_k \in \C^{M}$ be the channel of the direct link between the BS and UE $k$, $\HH_\ell\in\C^{M\times N}$ be the channel between the BS and RIS $\ell$, and $\z_{\ell k} \in \C^{N}$ be the channel between RIS $\ell$ and UE $k$.
		The indirect link that connects the BS and UE $k$ via RIS $\ell$ is the cascaded channel constructed by $\HH_\ell$ and $\z_{\ell k}$ \cite{chien21TWC}, i.e., $\HH_\ell \PHI_\ell \z_{\ell k}$. 
		
		We consider the cases that there are possible LoS paths among the BS, RIS, and UEs. Therefore, $\g_k$, $\HH_\ell$, and $\z_{\ell k}$ with both LoS and non-LoS paths are modeled using the Rician distribution. 
		% \begin{align}
			%     \g_k &= \sqrt{\beta_{0,k}}\Bigg(\sqrt{\frac{K_{0,k}}{K_{0,k}+1}} \bar{\g}_k + \sqrt{\frac{1}{K_{0,k}+1}} \tilde{\g}_k\Bigg)
			%     \\
			%     \HH_\ell &= \sqrt{\beta_{1,\ell}}\Bigg(\sqrt{\frac{K_{1,\ell}}{K_{1,\ell}+1}} \bar{\HH}_\ell + \sqrt{\frac{1}{K_{1,\ell}+1}} \widetilde{\HH}_\ell\Bigg)
			%     \\
			%     \z_{\ell k} &=\sqrt{\beta_{2,\ell k}}\Bigg( \sqrt{\frac{K_{2,\ell k}}{K_{2,\ell k}+1}} \bar{\z}_{\ell k} + \sqrt{\frac{1}{K_{2,\ell k}+1}} \tilde{\z}_{\ell k} \Bigg),
			% \end{align}
		In particular, for notation simplicity, channels $\g_k$, $\HH_\ell$, and $\z_{\ell k}$ can be rewritten as 
		\begin{align}\label{eq:gk}
			\g_k &= \sqrt{\beta_{0,k}^{\text{LoS}}} \bar{\g}_k + \sqrt{\beta_{0,k}^{\text{NLoS}}} \tilde{\g}_k,
			\\ \label{eq:Hll}
			\HH_\ell &= \sqrt{\beta_{1,\ell}^{\text{LoS}}}\bar{\HH}_\ell + \sqrt{\beta_{1,\ell}^{\text{NLoS}}}\widetilde{\HH}_\ell,
			\\ \label{eq:zlk}
			\z_{\ell k} &=\sqrt{\beta_{2,\ell k}^{\text{LoS}}}\bar{\z}_{\ell k} + \sqrt{\beta_{2,\ell k}^{\text{NLoS}}} \tilde{\z}_{\ell k},
		\end{align}
		where $\beta_{0,k}^{\text{LoS}} \triangleq  \frac{\beta_{0,k}K_{0,k}}{K_{0,k}+1}$, $\beta_{0,k}^{\text{NLoS}} \triangleq  \frac{\beta_{0,k}}{K_{0,k}+1}$, $\beta_{1,\ell}^{\text{LoS}} \triangleq  \frac{\beta_{1,\ell}K_{1,\ell}}{K_{1,\ell}+1}$, $\beta_{1,\ell}^{\text{NLoS}} \triangleq  \frac{\beta_{1,\ell}}{K_{1,\ell}+1}$, $\beta_{2,\ell k}^{\text{LoS}} \triangleq  \frac{\beta_{2,\ell k}K_{2,\ell k}}{K_{2,\ell k}+1}$, and $\beta_{2,\ell k}^{\text{NLoS}} \triangleq  \frac{\beta_{2,\ell k}}{K_{2,\ell k}+1}$. Here, $K_{0,k}, K_{1,\ell}$, and $K_{2,\ell k}$ are the Rician $K$-factors; $\beta_{0,k}, \beta_{1,\ell}$, and $\beta_{2,\ell k}$ are the large-scale fading coefficients; $\bar{\g}_k, \bar{\HH}_\ell$, and $\bar{\z}_{\ell k}$ represent the LoS components of the corresponding channels, while 
		$\tilde{\g}_k, \widetilde{\HH}_\ell$, and $\tilde{\z}_{\ell k}$ represent the non-line-of-sight components. The elements of $\tilde{\g}_k, \widetilde{\HH}_\ell$, and $\tilde{\z}_{\ell k}$ are independently and identically distributed (i.i.d.) complex Gaussian random variables with zero mean and unit variance.

		\vspace{-2mm}
		\subsubsection{Channel Estimation at the BS with the Uplink Pilot Training}
		To prepare for the signal processing of the downlink transmission, the BS needs to acquire  knowledge of the channels in each coherence interval via channel estimation. The channels are independently estimated from pilot sequences of length $\tau_p$ transmitted simultaneously by all the UEs. Denote by $\sqrt{\tau_p}\VARPHI_k$ the pilot sequence of UE $k$, where $\VARPHI_k\in\C^{\tau_p}$ and $\|\VARPHI_k\|^2 = 1$. The pilot sequences are assumed to be mutually orthogonal, i.e., $\VARPHI_k^H\VARPHI_k = 1$, and $\VARPHI_k^H\VARPHI_{k'}=0, \forall k\neq k'$, where $(.)^H$ represents the conjugate transpose operator. The received pilot signals at the BS, $\Y_p \in \C^{M \times \tau_p}$, can be given as
		\begin{align}
			\nonumber
			\Y_p \!=\! \sum_{k\in\K}\sqrt{\rho\tau_p} \g_k \VARPHI_k^H + \sum_{k\in\K}\sum_{\ell\in\LL} \sqrt{\rho\tau_p} \HH_\ell \PHI_\ell \z_{\ell k} \VARPHI_k^H + \W_p,
		\end{align}
		where $\rho$ is is the normalized signal-to-noise ratio (SNR) of each pilot symbol, $\W_p \in \C^{M \times \tau_p}$ is the additive noise at the BS with i.i.d.  $\CN(0,1)$ elements. 
		
		Now, to estimate the desired channels from UE $k$, the received pilot signal $\Y_p$ is projected on $\VARPHI_k$ as
		\begin{align}
			\label{eq:ypk}
			\nonumber
			\y_{p,k} = 
			&\Y_p \VARPHI_k
			% \sum_{k\in\K}\sqrt{\rho\tau_p} \bigg( \g_k + \sum_{\ell\in\LL}\HH_\ell \PHI_\ell \z_{\ell k} \bigg) 
			\\
			% &+ 
			= & 
			% \sum_{k\in\K}
			\sqrt{\rho\tau_p} \bigg( \g_{k} +  \sum_{\ell\in\LL}\HH_\ell \PHI_\ell \z_{\ell k} \bigg) + \W_p\VARPHI_k,
		\end{align}
		where $\W_p\VARPHI_k \sim \CN (\pmb{0},\pmb{I}_M)$. Denote by 
		
		\vspace{-5mm}
		\begin{align}
			\label{eq:uk}
			\uu_{k} = \g_k + \sum_{\ell\in\LL} \HH_\ell \PHI_\ell \z_{\ell k}
		\end{align}
		the aggregated channel between the BS and UE $k$. Since $\uu_{k}$ comprises the direct channel $\g_k$ and the indirect channel $\sum_{\ell\in\LL}\HH_\ell \PHI_\ell \z_{\ell k}$ which is a sum of cascaded structures, it is nontrivial to apply minimum mean-square error (MMSE) estimation method to estimate separately $\g_k$, $\HH_\ell$, and $\z_{\ell k}$ as reported in \cite{chien20TWC}. Instead, in this work, we assume that the BS employs the linear MMSE estimation technique to estimate the aggregated channel $\uu_{k}$. 
		% The estimate channel $\hat{\uu}_{k}$ of $\uu_{k}$ can be calculated as \cite[Eq. (12.6)]{kay93}
		% \begin{align}
			%     % \hat{\uu}_{k} = \EEE\{\uu_k\} + \EEE \{\uu_{k}\y_{p,k}^H\} (\EEE \{\y_{p,k}\y_{p,k}^H\})^{-1}(\y_{p,k} - \EEE\{\y_{p,k}\}.
			%     \hat{\uu}_{k} = \EEE\{\uu_k\} + \CC_{\uu_k,\y_{p,k}} \CC_{\y_{p,k},\y_{p,k}}^{-1}(\y_{p,k} - \EEE\{\y_{p,k}\}).
			% \end{align}
		% Here, $\CC_{\x,\y}$ and $\CC_{\x,\x}$ is the cross-covariance matrix of $\x,\y$, and the covariance matrix of $\x$, respectively. $\EEE\{x\}$ is the expectation of random variable $x$.
		\vspace{-2mm}
		\begin{proposition}\label{pro1}
			For given phase shift matrices, the linear MMSE channel estimate $\hat{\uu}_{k}$ of $\uu_k$ is given by
			\begin{align} \label{eq:MMSEestimate}
				\hat{\uu}_{k} = \pmb{\mu}_k + \CC_{k}\left(\sqrt{\rho\tau_p}\CC_{k}  + \pmb{I}_M \right)^{-1}\left(\y_{p,k} - \sqrt{\rho\tau_p} \pmb{\mu}_k\} \right),
			\end{align}
			where 
			\begin{align} \label{eq:muk}
				\pmb{\mu}_k &= \sqrt{\beta_{0,k}^{\text{LoS}}} \bar{\g}_k + \sum_{\ell\in\LL} \sqrt{\beta_{1,\ell}^{\text{LoS}}}\bar{\HH}_\ell \PHI_\ell \sqrt{\beta_{2,\ell k}^{\text{LoS}}}\bar{\z}_{\ell k},\\
				\CC_{k} &= \sqrt{\rho\tau_p}\left(\beta_{0,k}^{\text{NLoS}} \pmb{I}_M  + \sum_{\ell\in\LL}  {\beta_{1,\ell}^{\text{LoS}}} \beta_{2,\ell k}^{\text{NLoS}}\bar{\HH}_\ell  \PHI_\ell \PHI_\ell^H \bar{\HH}_\ell^H \right.\nonumber\\
				&\hspace{-0.7cm}\left. + \sum_{\ell\in\LL} \beta_{1,\ell}^{\text{NLoS}} \left( \beta_{2,\ell k}^{\text{LoS}}\|\PHI_\ell \bar{\z}_{\ell k} \|^2 + \beta_{2,\ell k}^{\text{NLoS}}\tr\{ \PHI_\ell \PHI_\ell^H\right) \pmb{I}_M  \right),
			\end{align}
			where $\tr(\X)$ is the trace of a matrix $\X$, $\|\x\|$ is the $2$-norm of a vector $\x$.
			\vspace{-3mm}
			\begin{proof}
				See Appendix.
				% ~\ref{proof_prop1}.
			\end{proof}
		\end{proposition}

		% In the next section, we will propose approaches to estimate the aggregated channels for given the matrices of phase shifts $\PHI_\ell, \forall \ell$. 
		\vspace{-3mm}
		\subsubsection{Downlink Data Transmission}
		By using channel reciprocity, the BS first treats the channel estimates obtained in the uplink training phase as the true channels to perform linear precoding on message symbols, and then sends the precoded signals to the UEs using the remaining $\tau_c-\tau_p$ transmission symbols per coherence interval. 
		Let $s_k(n)$ be the symbol intended for UE $k$, and $n \in \{1,\dots,\tau_c-\tau_p\}$ be the index of message symbol in each coherence interval, where $\EEE \{|s_k(n)|^2\}=1$, $\EEE \{.\}$ presents the expectation operator. Assuming a linear precoding vector $\aaa_k \in\C^M$ is applied to every UE $k$, the signal that the BS sends to all the UEs is
		\begin{align}
			\x(n) = \sum_{k\in\K} \sqrt{\rho_d\eta_k} \aaa_k s_{k}(n),
		\end{align}
		where $\rho_d$ is the maximum downlink transmit power and $\eta_k$ is a power control coefficient chosen to satisfy $\EEE \{\|\x(n)\|^2\}\leq \rho_d$. 
		
		The received signal $y_k(n)$ at UE $k$ is
		\begin{align}
			\label{yk}
			\nonumber
			y_k(n) &= \uu_k^H \x(n) + \tilde{w}_k(n) = \sqrt{\rho_d\eta_k} \uu_k^H \aaa_k s_{k}(n) 
			\\
			\nonumber
			&\quad+ \sum_{k'\in\K \setminus \{k\}} \sqrt{\rho_d\eta_{k'}} \uu_k^H \aaa_{k'} s_{k'}(n) + \tilde{w}_k(n)
			\\
			& = \alpha_{kk} s_{k}(n) +  \sum_{k'\in\K \setminus \{k\}} \alpha_{kk'} s_{k'}(n) + \tilde{w}_k(n),
		\end{align}
		where $\tilde{w}_k(n) \sim \CN(0,1) $ is the additive noise, and 
		\begin{align}
			\alpha_{kk'} \triangleq \sqrt{\rho_d\eta_{k'}} \uu_k^H \aaa_{k'},
		\end{align}
		% $\alpha_{kk'} \triangleq \sqrt{\rho_d\eta_{k'}} \uu_k^H \aaa_{k'}$ 
		is the \textit{effective channel gain}. The first term of \eqref{yk} is the \textit{desired signal} of UE $k$, while the remaining constructs an \textit{interference-plus-noise} term. 
		% Then, $s_k(n)$ is detected based on the observation of $y_k(n)$.
		
		% \subsection{Downlink Channel Estimation at UEs}
		\vspace{-2mm}
		\section{Effective Channel Estimation at UEs: \\Harding Bound and Model-Based Solutions}
		\vspace{-1mm}
		\label{sec:model-based}
		To detect $s_k(n)$ based on $y_k(n)$, UE $k$ needs to know the effective channel gain $\alpha_{kk}$ in the desired signal and the average power of the interference-plus-noise term in \eqref{yk}. Since $\alpha_{kk}$ varies in each coherence interval, estimating $\alpha_{kk}$ at every UE $k$ is very challenging, which calls for an efficient downlink channel estimation solution. One possible solution is to let the BS transmit downlink pilots to the UEs. However, with this solution, when the number of UEs is large, the downlink pilots take a large part of the coherence interval, which consequently reduces the spectral efficiency significantly \cite{hien17TWC}. Thus, channel estimation techniques without any downlink pilots are preferable. 
		
		\subsubsection{Hardening Bound}
		With the Hardening Bound scheme,  $\alpha_{kk}$ can be estimated by $\EEE \{\alpha_{kk}\}$, which is the nominal approach in massive MIMO \cite{hien17TWC}. The reason behind is that in massive MIMO systems, the channels harden thanks to the large number of antennas, i.e., $\alpha_{kk} \approx \EEE \{\alpha_{kk}\}$. However, the channels in RIS-assisted massive MIMO systems suffer from extra random fluctuations from the indirect links, such that the channel hardening property is not as strong as that in massive MIMO systems. Therefore, using $\EEE \{\alpha_{kk}\}$ as the true $\alpha_{kk}$ may result in a poor estimation performance. Based on these observations, in the following, we introduce an alternative solution named Model-Based that improves the estimation performance of RIS-assisted massive MIMO systems without using downlink pilots.

		\subsubsection{Model-Based}
		The model-based solution is based on the blind channel estimation approach in \cite{hien17TWC}. In our work, we extend \cite{hien17TWC} from massive MIMO to RIS-assisted massive MIMO systems. The key idea is to exploit the asymptotic properties of the sample mean power of the received signals in each coherence interval for channel estimation. Specifically, UE $k$ can compute the sample mean of the received signals in the current coherence interval as
		\begin{align}
			\label{xik}
			\xi_k = \frac{\sum_{n=1}^{\tau_c-\tau_p}|y_k(n)|^2}{\tau_c-\tau_p}.
		\end{align}
		By using \eqref{yk} and the law of large numbers, we have, as $\tau_c \rightarrow \infty$ for a fixed $\tau_p$, 
		\begin{align}
			\xi_k - \Bigg( |\alpha_{kk}|^2 +  \sum_{k'\in\K \setminus k} |\alpha_{kk'}|^2 + 1\Bigg) \overset{P}{\rightarrow} 0,
		\end{align}
		where $\overset{P}{\rightarrow}$ denotes convergence in probability. 
		On the other hand, following the law of large numbers, $\sum_{k'\in\K \setminus k} |\alpha_{kk'}|^2$ can be approximated by its mean because it is a sum of many terms. 
		Therefore, when $\tau_c$ and $K$ are large, 
		\begin{align}
			\label{xik:approx}
			\xi_k \approx |\alpha_{kk}|^2 + \delta_k,
		\end{align}
		where 
		% $\delta \triangleq \EEE \left\{ \sum_{k\in\K \setminus k'} |\alpha_{kk'}|^2 \right\} + 1$.
		\begin{align}
			\label{delta}
			\delta_k \triangleq \EEE \Bigg\{ \sum_{k'\in\K \setminus k} |\alpha_{kk'}|^2 \Bigg\} + 1,
		\end{align}
		is the mean power of the interference-plus-noise term. 
		% Importantly, the approximation \eqref{xik:approx} is expected to perform well even with a small value of $K$. This is because when $K$ is small, $|\alpha_{kk'}|^2 \ll |\alpha_{kk}|^2$ with high probability, then $\sum_{k\in\K \setminus k'} |\alpha_{kk'}|^2 \ll |\alpha_{kk}|^2$ with high probability, which means $\sum_{k\in\K \setminus k'} |\alpha_{kk'}|^2$ can be approximated by its mean. 
		
		Now, from the approximation \eqref{xik:approx}, we can estimate $\alpha_{kk}$ via $\xi_k$ as
		\begin{align}
			\alpha_{kk} \approx \sqrt{\xi_k - \delta_k},
		\end{align}
		In the case of non-positive value of $\xi_k - \delta_k$, we estimate $\alpha_{kk} \approx \EEE \{\alpha_{kk}\}$ as the Hardening Bound solution. In other words, the estimate $\hat{\alpha}_{kk}$ of the effective channel $\alpha_{kk}$ is given by
		\begin{align}\label{a}
			\hat{\alpha}_{kk} =
			\begin{cases}
				\sqrt{\xi_k - \delta_k},& \xi_k > \delta_k \\
				\EEE \{\alpha_{kk}\}, & \mbox{otherwise}.
			\end{cases}
		\end{align}
		The disadvantage of the model-based solution is that it relies on asymptotic properties that are only reasonable when $\tau_c$ and $K$ are sufficiently large. Thus, there is no guarantee that this solution can perform well when $\tau_c$ or $K$ is small. This observation motivated us to design a novel solution that uses neural networks for effective channel estimation and tackle the disadvantage of the model-based solution. 
		
		\vspace{-2mm}
		\section{Effective Channel Estimation at UEs: Proposed Learning-Based Solution}
		\vspace{-2mm}
		\label{sec:learning-based}
		The key idea of our learning-based solution is to use a DNN to identify a proper mapping between the features of the UE's received signal and the effective downlink channel gain. In particular, the input vector $\UPSI_k$ of the neural network at UE $k$ has $4$ features: (i) the sample mean of the received signals $\xi_k$ in \eqref{xik}, (ii) the mean power of interference-plus-noise $\delta_k$ in \eqref{delta}, (iii) $\rho_d\eta_k\EEE\{\|\uu_k\|^2\}$, and (iv) $\EEE\{\alpha_{kk}\}$. 
		Here, the first, second, and fourth features are obtained similarly as in the Hardening Bound and Model-Based approaches. The output of the DNN model is $\vartheta_k$, which is expected to be the effective downlink gain $\alpha_{kk}$. Note that there are other features that can be used as the inputs to the neural network. However, we aim to keep the number of input features small for ease of implementation. We select the set of these four features because it provides the best performance during our feature selection process. 
		
		\begin{table}[!t]
			\renewcommand{\arraystretch}{1.0}
			\caption{Proposed Deep Neural Network Layout}
			\label{NN:design}
			\begin{center}
				\vspace{-4mm}
				\begin{tabular}{|c|c|c|c|}
					%\hline
					\hline
					\textbf{}  & \textbf{Neurons} & \textbf{Parameters} & \textbf{Activation function}  
					\\
					\hline
					Layer 1      & $32$ & $160$ & ReLU \\
					\hline
					Layer 2      & $64$ & $2112$ & ReLU  \\
					\hline
					Layer 3      & $128$ & $8320$ & ReLU  \\
					\hline
					Layer 4      & $64$ & $8256$ & ReLU  \\
					\hline
				\end{tabular}
			\end{center}
			\vspace{-2mm}
		\end{table}
		
		\subsubsection{Proposed Deep Neural Network Design}
		
		To make a mapping from $\UPSI_k$ to $\vartheta_k$,
		% \textcolor{red}{(@Tung: 1)This parameter has not defined yet; 2) what is the difference between  $\theta_k$ and $\hat{\theta}_k^d$ presented later; 3) You may want to use another notation since $\theta$ has been used to denote the phase shifts)}
		we consider a fully-connected feed-forward neural network with $Q=4$ hidden layers and the sizes of these layers are given in Table~\ref{NN:design}. Note that it is possible to have many different designs of neural network for a given prediction accuracy. In this work, we choose the  fully-connected neural network and its hyper-parameters by using the trial-and-error approach. Specifically, we increase the hyper-parameters gradually and try different activation functions. The current layout of the neural networks in Table~\ref{NN:design} offers good performance in terms of NMSE which is defined as
		\begin{align}
			\text{NMSE}_k \triangleq  
			% \frac{1}{K} \sum_{k\in\K}
			\frac{\EEE \left\{|\alpha_{kk}-\hat{\alpha}_{kk}|^2\right\}}{\EEE \left\{|\alpha_{kk}|^2\right\}} .
		\end{align}
		
		\subsubsection{Training Deep Neural Network for Channel Estimation}
		We exploit the supervised learning approach to train the proposed deep neural network as follows. First, we create a data training set $\{\UPSI_k^d,\hat{\vartheta}_k^d\}_{d=1}^D$, each data point of which includes the input vectors $\UPSI_k^d$ and the desired output $\hat{\vartheta}_k^d$. Here, $\hat{\vartheta}_k^d$ is the true effective channel gain $\alpha_{kk}$, $D$ is the size of the training set, and $d$ is the index of data point. Let $\Omega$ be the set of weights assigned to the proposed neural network, and $\tilde{\alpha}_{kk}$ the output obtained after training process. We also use the mean square error (MSE) for the loss function, i.e., 
		\begin{align}
			\LL_k(\Omega) \triangleq \EEE \{|\tilde{\alpha}_{kk}-\alpha_{kk}|^2 \}.
		\end{align}
		Then, we train the proposed neural network with a standard process that is discussed in \cite{goodfellow16}. The training process is executed offline. Once this process is completed, the proposed neural network is then used online. To adapt the specific continuous mapping in this paper, our proposed DNN has a different design from that in \cite{amin21TCOM} in terms of the input features, the number of layers and the parameters, and the type of loss functions.
		
		% Complexity = ?
		
		\subsubsection{Acquiring Data Set at UEs in the Offline Mode}
		Acquiring the data set $\{\UPSI_k^d,\hat{\vartheta}_k^d\}_{d=1}^D$ at each UE $k$ is the key of the training process. First, UE $k$ can obtain $\xi_k$ by taking the sample mean of its received signals via \eqref{yk}. Since $\delta_k$, $\rho_d\eta_k\EEE\{\|\uu_k\|^2\}$, and $\EEE\{\alpha_{kk}\}$ depend on the known statistical properties of the channels, they can be easily obtained by UE $k$. 
		% Then, $\delta_k$ is obtained by taking the mean power of the interference-plus-noise term. 
		The main challenge is to obtain the true effective channel gain $\alpha_{kk}$ for the desired output $\hat{\vartheta}_k^d$. Fortunately, thanks to the offline training process, the BS can use the remaining coherence interval $\tau_c - \tau_p$ for beamforming orthogonal pilot sequences to the UEs. Since $\tau_c$ is normally large, UE $k$ can estimate
		% $\uu_k$, and hence, 
		$\alpha_{kk}$  accurately. 
		% Therefore, the UEs can compute $\rho_d\eta_k\EEE\{\|\uu_k\|^2\}$, and then send both $\rho_d\eta_k\EEE\{\|\uu_k\|^2\}$ and $\alpha_{kk}$ to each UE $k$.

		\vspace{-2mm}
		\section{Numerical Examples}
		\vspace{-1mm}
		\label{sec:sim}
		\subsubsection{Network Setup and Parameters Setting}
		\vspace{-0mm}
		We consider a scenario that a BS 
		% with $M=50$ antenna 
		is located at $(0,0)$. $L=2$ RISs are located at $(10,30)$
		% , $(10, 10)$, $(10,-10)$, 
		and $(10,-30)$ to ensure LoS paths from the BS.
		$K=10$ UEs are randomly located in the $100$ m $\times 100$ m area that extends from $(150, -50)$ and $(250,50)$. The BS and RISs are assumed to have the same altitude above the UEs by $10$ m. The existence of LoS paths for the channels from the BS or the RISs to the UEs is modeled in a probabilistic manner and the Rician $K$-factors  are given in \cite[Sec. 5.5.3]{3GPP20} and \cite{ozdogan19TCOM}. The large-scale fading coefficients in both cases that the LoS paths exist and do not exist are  given in (68) and (69) of \cite{ozdogan19TCOM}, respectively. 
		% $\beta\text{[dB]} = -30.18 -26\log_{10}(d) + F$, where $d$ is the distance between two objectives and $F\sim \NN(0,\sigma)$.

		The BS is equipped with a uniform linear array (ULA) located on the $y$-axis with antenna spacing $d_{B} = \lambda/2$, where $\lambda$ denotes the wavelength, while the RISs are equipped with uniform planar array (UPA) located parallel to the $x$-$z$ plane with RIS element spacing $d_R=\lambda/4$.
		The LoS components $\bar\g_k$, $\bar\HH_\ell$, and $\bar\z_{\ell k}$ are assumed to be known and given by 
		\cite{zhang20JSAC}
		\begin{align}
			&\bar\g_k = \aaa_B(\theta_{BU,k}^D)
			\\
			&\bar\HH_\ell = \aaa_B(\theta_{BR,\ell}^D)\aaa_R^H(\theta_{BR,\ell}^A,\psi_{BR,\ell}^A)
			\\
			&\bar\z_{\ell k} = \aaa_R(\theta_{RU,\ell k}^D,\psi_{RU, \ell k}^D).
		\end{align}
		Here, $\aaa_B(\theta)\in\C^{M}$ is the array response of the ULA of the BS, where $[\aaa_B(\theta)]_m= e^{j2\pi(m-1)d_B\sin \theta/\lambda}$ and $\theta \in [0,2\pi)$. $\theta_{BR,\ell}^D$ is the angle-of-departure (AoD) from the BS to RIS $\ell$; $\aaa_R(\theta,\psi)\in\C^N$ is the array response of the UPA of RIS, where $[\aaa_R(\theta,\psi)]_n = e^{j2\pi d_R (\lfloor\frac{n}{N_x}\rfloor \sin \psi \sin \theta + (n-\lfloor\frac{n}{N_x}\rfloor N_x)\sin\psi \cos \theta)/\lambda}$, $N_x = \min (N,5)$ denotes the number of RIS elements in each row along the $x$-axis, $\theta \in [0,2\pi)$, $\psi \in [-\pi/2,\pi/2)$ and $\lfloor x \rfloor$ denotes the closest integer no larger than $x$;  $(\theta_{BR,\ell}^A,\psi_{BR,\ell}^A)$ are the azimulth and elevation angle-of-arrival from the BS to RIS $\ell$. $(\theta_{RU,\ell k}^D,\psi_{RU, \ell k}^D)$ are the azimuth and elevation AoD from RIS $\ell$ to UE $k$. 
		% We use the 3GPP model \cite{3GPP20} for $\g_k$, $\HH_\ell$, and $\z_{\ell k}$. For example, the probability of a LoS path between the BS and UE $k$ is
		% \begin{align}
			%     \Pr(\text{LoS}) = 
			%     \begin{cases}
				%       \frac{300-d_{0,k}}{300}, & 0\leq d_{0,k} \leq  300 \text{ m}\\
				%       0, & d_{0,k} > 300 \text{ m},
				%     \end{cases}
			% \end{align}
		% where $d_{0,k}$ is the distance between the BS and UE $k$. 
		
		% The values of the Rician factors and the large-scale path-loss coefficients are calculated as in \cite{ozdogan19TCOM}. 
		
		The phase shift applied at element $n$ of RIS $\ell$ is modeled as $\nu_{\ell,n}=a_{\ell,n}(\theta_{\ell,n}) e^{j\theta_{\ell,n}}$, where $\theta_{\ell,n} \in [-\pi,\pi)$,  $a_{\ell,n}(\theta_{\ell,n}) = (1-a_{\min})(\frac{\sin (\theta_{\ell,n}-\phi)+1}{2})^b+a_{\min}$ is the amplitude, and $a_{\min}\geq 0, \phi \geq 0, b \geq 0$ are the constants related to the specific circuit implementation \cite{samith20TCOM}. The phase shift model parameters are set as: $a_{\min} = 0.2, b = 1.6, \phi = 0.43\pi$ \cite{samith20TCOM}, and $\theta_{\ell,n}$ is uniformly chosen in $[-\pi,\pi)$. We assume the BS uses the maximum-ratio precoding and equal power allocation schemes. Specially, the precoder and power control coefficient for UE $k$ are
		\begin{align}
			\aaa_k = \frac{\hat{\uu}_k}{\sqrt{\EEE\{\|\hat{\uu}_k\|^2\}}},
			\eta_k = \frac{1}{K}, 
		\end{align}
		where the normalization term guarantees $\EEE\{\|\aaa_k\|^2\}=1$. We use QPSK modulation,  noise power of $\sigma_0^2\!=\!-92$ dBm, and $\tau_c=500$. The uplink pilot power is $0.1$ W and the downlink power at the BS is $10$ W. 
		% where $\vv_k \in \C^{M}$ is the $k$-th column of the pseudo-inverse matrix $\widehat{\U}(\widehat{\U}^H\widehat{\U})^{-1}$ and $\widehat{\U}\triangleq [\hat{\uu}_1,\dots,\hat{\uu}_K] \in \C^{M\times K}$.
		
		Since the locations of UEs and shadow fading coefficients are identical and independently distributed, a large training data set at each UE $k$ becomes typical, i.e., can be utilized for the same neural network of every UE. Based on this observation, we train our neural network for only a typical UE instead of all $K$ UEs. In particular, we generate a data set of $D=2\times 10^6$ samples for $2000$ large-scale realizations and $1000$ small-scale fading realizations for each large-scale realization. Here, we choose $400000$ samples for training, $100000$ for validation, and $1500000$ for testing. The neural network is implemented using The Keras open-source library in the environment of Python $2.9$ on an Intel I7-10750H processor. For the training phase, we use Adam optimizer \cite{kingma15ICLR} with a learning rate of $0.01$, $200$ epochs, and a batch size of $128$. 
		Each sample mean is taken over $1000$ small-scale realizations. 
		
		\begin{figure}[t!]
			\centering
			\vspace{-3mm}
			{\includegraphics[width=0.36\textwidth]{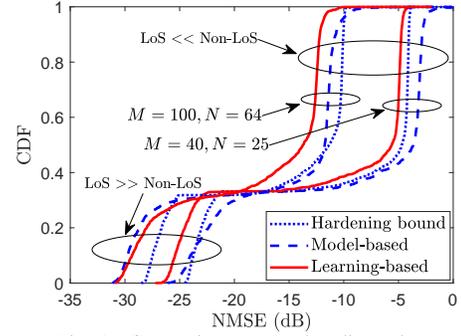}\label{fig:a}}
			\vspace{-4mm}
			\caption{Comparison among baseline schemes.}
			\label{Fig1}
		\end{figure}
		
		\vspace{-1mm}
		\subsubsection{Results and Discussions}
		% \vspace{-1mm}
		Fig.~\ref{Fig1} compares the state-of-the-art approaches, i.e., \textbf{Hardening bound} and \textbf{Model-based}, with the proposed \textbf{Learning-based} approach in terms of NMSE. As seen, the proposed learning-based outperform the Hardening Bound and Model-Based approaches in the cases of $(M=40,N=25)$ and $(M=100,N=64)$ with LoS paths dominated by non-LoS paths. This is because in these cases, the level of channel hardening is low, the Hardening Bound and Model-Based approaches perform poorly, while the learning-based approach with the additional knowledge of the input features obtained from these two approaches offers noticeable improvements. On the other hand, when LoS paths dominate non-LoS paths, the channel gains are considered to be approximately deterministic. Both the Hardening Bound and Model-Based are sufficient for channel estimation with $\text{NMSE} \leq -20$ dB.

\begin{figure}[t!]
			\centering
			\vspace{-0mm}
			{\includegraphics[width=0.36\textwidth]{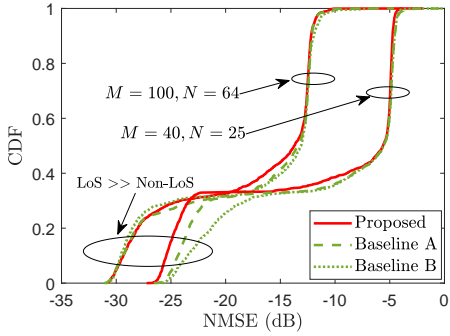}\label{fig:a}}
			\vspace{-3mm}
			\caption{Learning-based approaches with different inputs.}
			\label{Fig2}
			\vspace{-5mm}
		\end{figure}

		To justify for why we choose $4$ input features discussed in Sec.~\ref{sec:learning-based}, Fig.~\ref{Fig2} compares our \textbf{Proposed} Learning-based with its two versions: \textbf{Baseline A} using $3$ features (i), (ii), (iii), and \textbf{Baseline B} using $3$ features (i), (ii), (iv). As seen, although $3$ input features are enough for the case of large values of $M$ and $N$, they seem not to be sufficient for the case of small values of $M$ and $N$, especially when LoS paths dominate the non-LoS paths. Importantly, the training and running times of our learning-based approach with 4 input features are not much larger than the baselines with 3 input features as seen in Table~\ref{Tab:Time}. Note that the running time by more powerful computing units in real scenarios can be much smaller that those in Table~\ref{Tab:Time}.

		\begin{table}[!t]
			\renewcommand{\arraystretch}{1.0}
			\caption{Learning-based scheme with different inputs}
			\label{Tab:Time}
			\begin{center}
				\vspace{-4mm}
				\begin{tabular}{|c|c|c|c|}
					%\hline
					\hline
					\textbf{} & \textbf{Baseline A} & \textbf{Baseline B} & \textbf{Proposed}  
					\\
					\hline
					Training time & $339.951$ s & $343.6910$ s & $366.6387$ s \\
					\hline
					Running time & $0.0891$ s & $0.0878$ s & $0.0951$ s \\
					\hline
				\end{tabular}
			\end{center}
			\vspace{-2mm}
		\end{table}
		% 2 figures. Option 1: one is SNR vs NMSE. one is $N$ vs NMSE. 
		% Option 2: CDF NMSE. one with $N = 8 \times 8 $. one with $N = 16 \times 16 $.
		
		%\begin{figure}[t!]
		%  \centering
		%  {\includegraphics[width=0.36\textwidth]{....eps}\label{fig:a}}
		%  \caption{....}
		%  \label{Fig:sim1}
		%  \vspace{-5mm}
		%\end{figure}

		\vspace{-3mm}
		\section{Conclusion}
		\label{sec:con}
		\vspace{-1mm}
		This work has studied the channel estimation of RIS-assisted downlink massive MIMO systems at both the BS and UE sides. We derived closed-form expressions for the MMSE channel estimates at the BS. For the channel estimation at the UEs, two state-of-the-art approaches, i.e., Hardening Bound and Model-Based, from the traditional massive MIMO were extended to use in the RIS-assisted massive MIMO systems. Then, a learning-based approach was proposed to improve the performance of these approaches in the cases of a low level of channel hardening, and confirmed by simulation results. Note that the phase shifts of RISs are fixed in this work but can be further optimized to improve the channel estimation accuracy in future works.

		\vspace{-3mm}
		\appendix
		\vspace{-1mm}
		% \appendices
		%  \section{Proof of Proposition~\ref{pro1} }\label{proof_prop1}
		By using  \cite[Eq. (12.6)]{kay93}, the linear MMSE estimate of $\uu_{k}$ can be calculated as
		\begin{align} \label{eq:hatukk_proof}
			% \hat{\uu}_{k} = \EEE\{\uu_k\} + \EEE \{\uu_{k}\y_{p,k}^H\} (\EEE \{\y_{p,k}\y_{p,k}^H\})^{-1}(\y_{p,k} - \EEE\{\y_{p,k}\}.
			\hat{\uu}_{k} = \EEE\{\uu_k\} + \CC_{\uu_k,\y_{p,k}} \CC_{\y_{p,k},\y_{p,k}}^{-1}(\y_{p,k} - \EEE\{\y_{p,k}\}),
		\end{align}
		where $\CC_{\x,\y}$ and $\CC_{\x,\x}$ is the cross-covariance matrix of $\x,\y$, and the covariance matrix of $\x$, respectively. %and $\EEE\{x\}$ is the expectation of random variable $x$.
		\begin{itemize}
			\item Compute $ \EEE\{\uu_k\}$: From \eqref{eq:uk}, we have
			\begin{align}\label{eq:Euk}
				&\EEE\{\uu_k\} 
				= \EEE\{\g_k\} + \sum_{\ell\in\LL} \EEE\{\HH_\ell \PHI_\ell \z_{\ell k}\}\nonumber\\
				&= \sqrt{\beta_{0,k}^{\text{LoS}}} \bar{\g}_k + \sum_{\ell\in\LL} \sqrt{\beta_{1,\ell}^{\text{LoS}}}\bar{\HH}_\ell \PHI_\ell \sqrt{\beta_{2,\ell k}^{\text{LoS}}}\bar{\z}_{\ell k}.
			\end{align}
			
			\item Compute $\EEE\{\y_{p,k}\}$: From \eqref{eq:ypk} and \eqref{eq:uk}, we obtain
			\begin{align}\label{eq:Eypk}
				\EEE\{\y_{p,k}\} = \sqrt{\rho\tau_p} \EEE\{\uu_k\}. 
			\end{align}
			
			\item Compute $\CC_{\uu_k,\y_{p,k}}$: 
			\begin{align}\label{eq:Cukyk}
				&\CC_{\uu_k,\y_{p,k}} 
				= \EEE\{\left( \uu_k - \EEE\{\uu_k\} \right)\left( \y_{p,k} - \EEE\{\y_{p,k}\} \right)^H \}\nonumber\\
				&= \sqrt{\rho\tau_p}\EEE\{\left( \uu_k - \EEE\{\uu_k\} \right)\left(  \uu_k - \EEE\{\uu_k\}\} \right)^H \}\nonumber\\
				&= \sqrt{\rho\tau_p} \Bigg(\beta_{0,k}^{\text{NLoS}} \pmb{I}_M  + \sum_{\ell\in\LL}  {\beta_{1,\ell}^{\text{LoS}}} \beta_{2,\ell k}^{\text{NLoS}}\bar{\HH}_\ell  \PHI_\ell \PHI_\ell^H \bar{\HH}_\ell^H \nonumber
				\\
				& + \sum_{\ell\in\LL} \beta_{1,\ell}^{\text{NLoS}} \left( \beta_{2,\ell k}^{\text{LoS}}\|\PHI_\ell \bar{\z}_{\ell k} \|^2 + \beta_{2,\ell k}^{\text{NLoS}}\tr\{ \PHI_\ell \PHI_\ell^H\right) \pmb{I}_M  \Bigg).
			\end{align}
			
			\item Compute $\CC_{\y_{p,k},\y_{p,k}}$: We have
			\begin{align}\label{eq:Cykyk}
				&\CC_{\y_{p,k},\y_{p,k}} 
				= \EEE\{\left( \y_{p,k} - \EEE\{\y_{p,k}\} \right)\left( \y_{p,k} - \EEE\{\y_{p,k}\} \right)^H \}\nonumber\\
				&= \sqrt{\rho\tau_p}\CC_{\uu_k,\y_{p,k}}  + \EEE\{\W_p\VARPHI_k \left( \y_{p,k} - \EEE\{\y_{p,k}\} \right)^H  \}  \nonumber\\
				&= \sqrt{\rho\tau_p}\CC_{\uu_k,\y_{p,k}}  + \pmb{I}_M  .
			\end{align}
		\end{itemize}
		Substituting \eqref{eq:Euk}--\eqref{eq:Cykyk} into \eqref{eq:hatukk_proof}, we arrive at \eqref{eq:MMSEestimate}.

		% you can choose not to have a title for an appendix
		% if you want by leaving the argument blank
		%\section{}

		% use section* for acknowledgement
		%\section*{Acknowledgment}
		%We would like to thank the reviewers for constructive comments.
		%This work is supported in part by an ECR-HDR scholarship from The University of Newcastle and in part by the Australian Research Council Discovery Project grant DP170100939.

		% Can use something like this to put references on a page
		% by themselves when using endfloat and the captionsoff option.
		\ifCLASSOPTIONcaptionsoff
		\newpage
		\fi

		\vspace{-2mm}
		\section*{Acknowledgment}
		\vspace{-2mm}
		The work of T. T. Vu and H. Q. Ngo was supported by the U.K. Research and Innovation Future Leaders Fellowships under Grant MR/S017666/1. The work of M. Matthaiou was supported by the European Research Council (ERC) under the European Union's Horizon 2020 research and innovation programme (grant agreement No. 101001331).
		
		\vspace{-2mm}
		\begin{spacing}{0.92}
			\bibliographystyle{IEEEtran}
			\bibliography{IEEEabrv,ref22}
		\end{spacing}
		
	\end{document}